\documentclass{PoS}

\usepackage[strict]{changepage}
\usepackage[?]{amsmath}
\usepackage{bbm}
\usepackage{psfrag}

     \title{Soft dynamics in heavy-light mesons}
\ShortTitle{On internal structure of the heavy-light mesons}

\author{%
Damir Be\'cirevi\'c, \speaker{Emmanuel~Chang}, Alain Le Yaouanc\\
Laboratoire de Physique Th\'eorique (B\^at.~210)~%
\footnote{Laboratoire de Physique Th\'eorique est une unit\'e mixte de recherche du CNRS, UMR 8627.}\\
Universit\'e Paris Sud, Centre d'Orsay,\\ 
F-91405 Orsay-Cedex, France\\
E-mail: \email{Emmanuel.Chang@th.u-psud.fr}
}

\abstract{%
We compute the radial distributions of the scalar, vector and axial charge density in the heavy-light mesons.  
We present the results obtained for both the lowest lying static heavy-light mesons as well as for their nearest excitations,
with $N_f=2$ dynamical quarks of Wilson (Clover) type. We used various improvements of the static heavy-quark actions.
From these distributions we were able to compute the corresponding charges and their radii  $\left< r^2\right>$, the results of which are also presented.
}

\FullConference{The XXVII International Symposium on Lattice Field Theory - LAT2009\\
		 July 26-31 2009\\
		 Peking University, Beijing, China}

\newcommand{\trout}[1]{\rule{0pt}{#1}}

\begin{document}


\section{Introduction}
Lattice QCD is the only ab initio method to compute non-perturbative properties of the bound states of quarks and gluons.
In this study, we focus on the dynamics of the light degrees of freedom inside a heavy-light meson. In the static limit
($m_Q \to \infty$), we can properly define and compute spatial distributions of \emph{matter}, of the electric charge and of the axial charge.
These can then be compared to various quark models which are still the main tool for describing the dynamics of excited states.
The spatial distributions  which we present here constitute a detailed testing ground for the quark models, for their choice of parameters, as well
as for the basic assumptions, and therefore may be a precious means to select among them and/or to improve them.
\section{Static heavy-quark propagator}
The  latticized version of the static heavy-quark effective theory Lagrangian (HQET), in the frame in which the heavy-quark is
at rest, reads~\cite{Eichten}
\[ {\cal L}_{\rm HQET}=\sum_x h^\dagger(x)\left[ h(x) - \mathbf{V_0(x-\hat 0)}^\dagger h(x-\hat 0)\right] + \delta m(g^2) h^\dagger(x)h(x) \]

\begin{adjustwidth}{-8pt}{0pt}
\begin{itemize}
\item The power divergent term $\delta m(g^2) h^\dagger(x)h(x)$ of the discretized HQET is a regularization artifact.
\item $V_0(x)$ is either the lattice gauge link variable in time direction, or a combination of links in the neighbourhood of the said link.
The former gives the Eichten-Hill action; with the latter, depending on the prescription, various actions are obtained, i.e.\ FAT6, HYP-1, HYP-2, \dots.
\end{itemize}
\end{adjustwidth}

\noindent The static heavy-quark propagator is given by
\[ S_{ h} (\vec x, 0; \vec y, t)= {1+\gamma_0\over 2} \delta(\vec x- \vec y)\ \theta(t) \prod_{n=0}^{t-1} V_0^\dagger (n) \]\,.
With the Eichten-Hill action, one immediately realizes that the signal of the two point static-light function deteriorates
beyond use long before the ground state of the heavy-light meson can be reliably extracted.

FAT6 action was the first proposed to ameriolate the situation. The more recent HYP-1 \cite{hasenfratz} and HYP-2 \cite{actions-dellamorte}
actions further reduce the noise to signal ratio (figure \ref{fig:1}). In this study, we show that HYP-1/2 can be improved by simply
iterating the HYP procedure.

In short, the HYP procedure consists in decorating a given temporal link by the links situated in its neighboring hyper-cube. 
We refer the reader to the references to HYP-1/2 above and our paper \cite{structure} for details. For earlier computations of these 
distributions see refs.~\cite{koponen1}.

\section*{Choice of the static action}
\begin{figure}[!hbt]
\begin{adjustwidth}{-1.575cm}{-1.575cm}
\begin{center}
\begin{tabular}{cc}
\psfrag{psfragTitle}{\Huge ${\cal E}_{eff}$}
\psfrag{psfragTAGA}{\Large\hspace{18pt} $\mathbf{FAT6}$}
\psfrag{psfragTAGB}{\Large\hspace{18pt} $\mathbf{HYP1^{\phantom{2}}}$}
\psfrag{psfragTAGC}{\Large\hspace{18pt} $\mathbf{HYP1^2}$}
\psfrag{psfragTAGD}{\Large\hspace{18pt} $\mathbf{HYP2^{\phantom{2}}}$}
\psfrag{psfragTAGE}{\Large\hspace{18pt} $\mathbf{HYP2^2}$}
\psfrag{psfragTAGF}{\Large\hspace{-8pt} $\mathbf{Eichten\ Hill}$}
\resizebox{77.575mm}{!}{\includegraphics{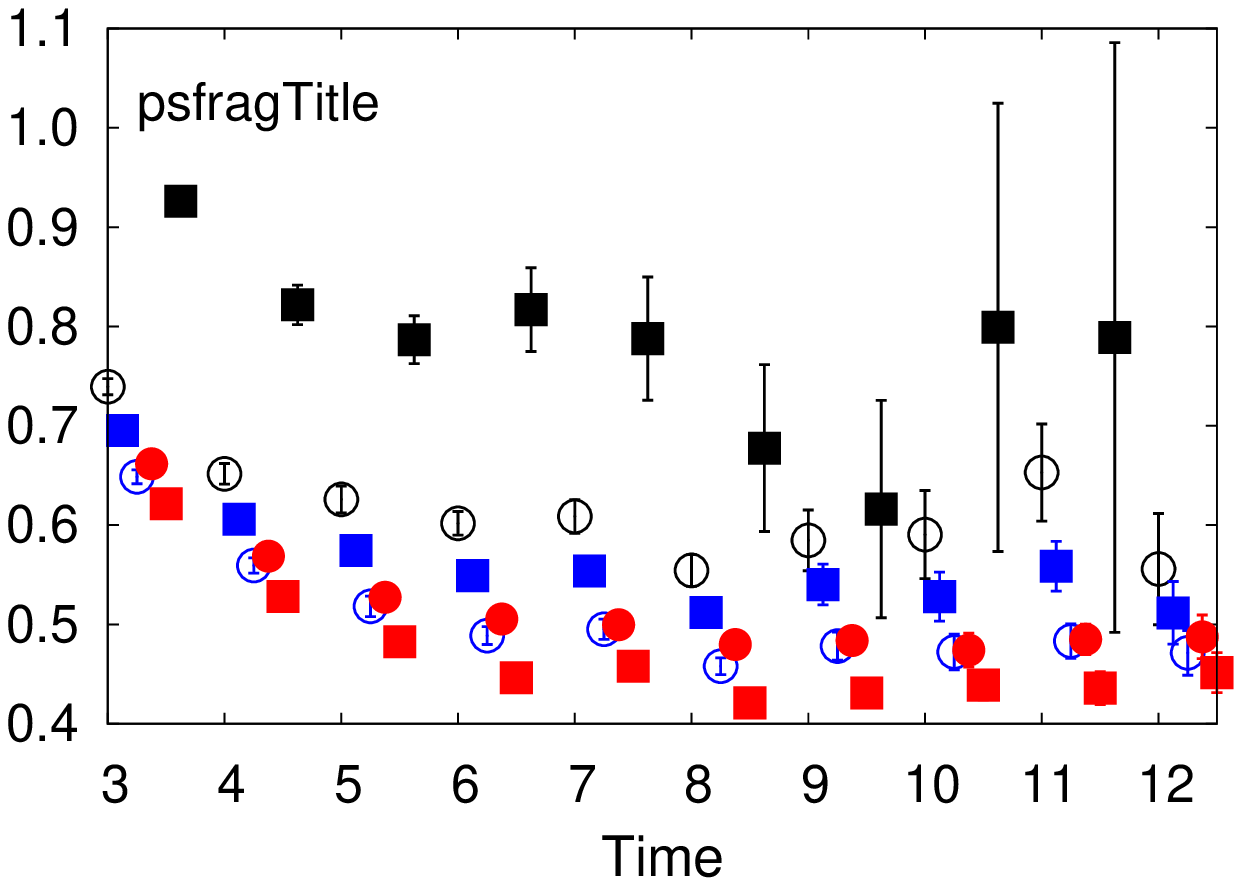}} &
\psfrag{psfragTitle}{\Huge $\frac{\mathbf{Error}}{\mathbf{Mean}}$}
\psfrag{psfragTAGA}{\Large\hspace{18pt} $\mathbf{FAT6}$}
\psfrag{psfragTAGB}{\Large\hspace{18pt} $\mathbf{HYP1}$}
\psfrag{psfragTAGC}{\Large\hspace{18pt} $\mathbf{HYP1^2}$}
\psfrag{psfragTAGD}{\Large\hspace{18pt} $\mathbf{HYP2}$}
\psfrag{psfragTAGE}{\Large\hspace{18pt} $\mathbf{HYP2^2}$}
\psfrag{psfragTAGF}{\Large\hspace{-8pt} $\mathbf{Eichten\ Hill}$}
\resizebox{77.625mm}{!}{\includegraphics{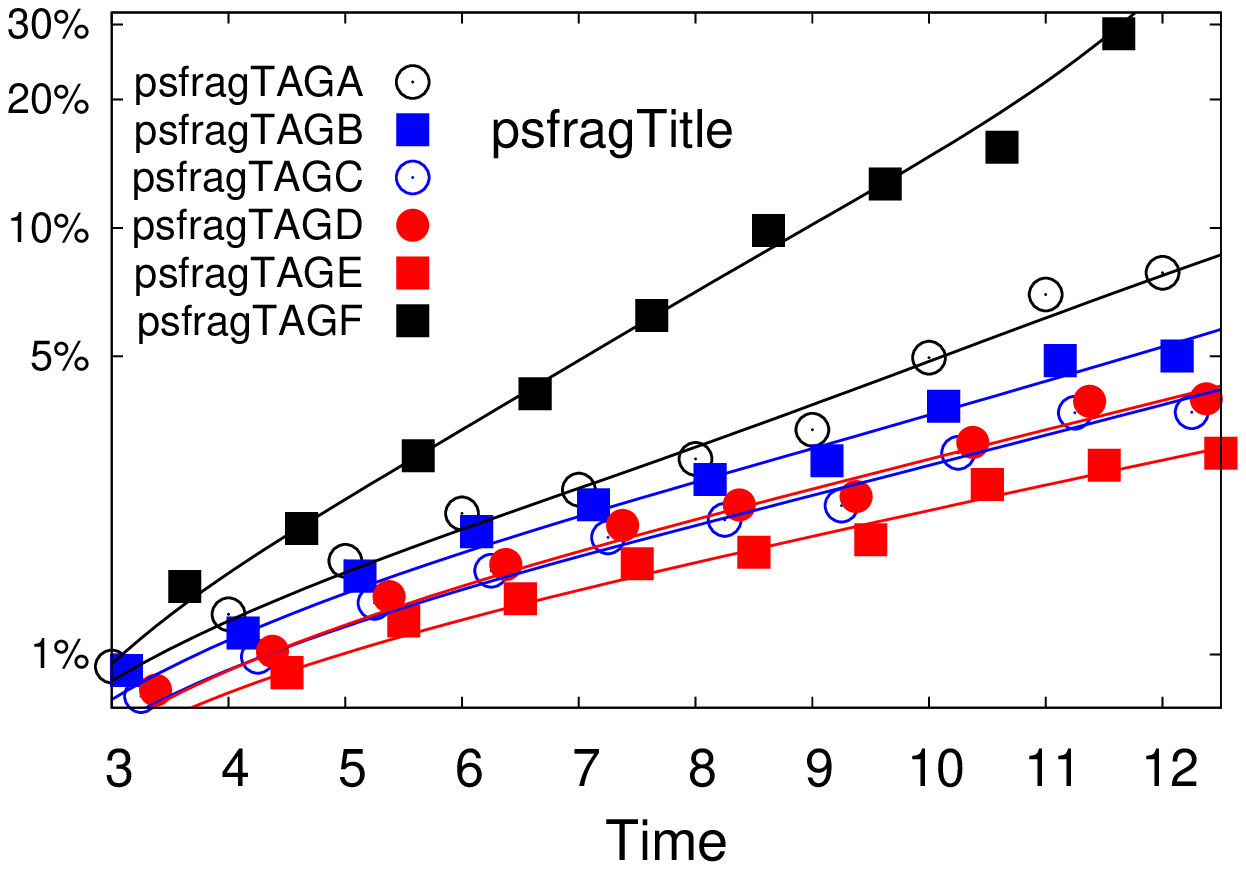}}
\end{tabular}
\end{center}
\end{adjustwidth}
\vspace{-16pt}
\begin{center}
\caption{\label{fig:1}
\footnotesize{\sl 
Plot on the left shows the effective mass obtained using six heavy-quark actions (light quark at $\kappa=0.1374$, which is somewhat
lighter but close to the physical strange quark mass). The right plot shows the (exponential) growth of the statistical error of the
two-point function as the time separation between the sources is increased\cite{exponential}.  Notice that the error on the correlation
function with HYP-2 heavy-quarks grows in the same way as with the HYP-1$^2$ one. \textbf{Both sink and source are smeared \`a la Boyle.}
}
} 
\end{center}
\vspace{-34pt}
\end{figure}

For the spatial distributions we consider, the heavy quark is only a spectator. It stands still at a point in space, while we
measure the values of our desired local operators at the light-light vertex as the light quark is pulled away from the fixed center.
This leaves us freedom in the choice of static quark action. Judging from figure \ref{fig:1}, HYP-2$^2$ action (HYP-2 iterated twice) is the best. Henceforth,
results presented are obtained using the that heavy quark action.
\addtolength\tabcolsep{-6pt}
\addtolength\tabcolsep{6pt}

Note also that smearing is highly important for the isolation of ground states in both two/three point function calculations.
In our study, we adopt the gauge invariant smearing of Boyle\cite{boyle}. For the parameters used, we refer the reader to our paper\cite{structure}.
A more detailed study on the effect of smearing will appear in another paper (work in progress).
\vspace{-8pt}
\section{Distributions}
\begin{figure}[!t]
\begin{adjustwidth}{-1.25cm}{-1.0cm}
\begin{center}
\begin{tabular}{ll}
{\resizebox{80mm}{!}{\includegraphics{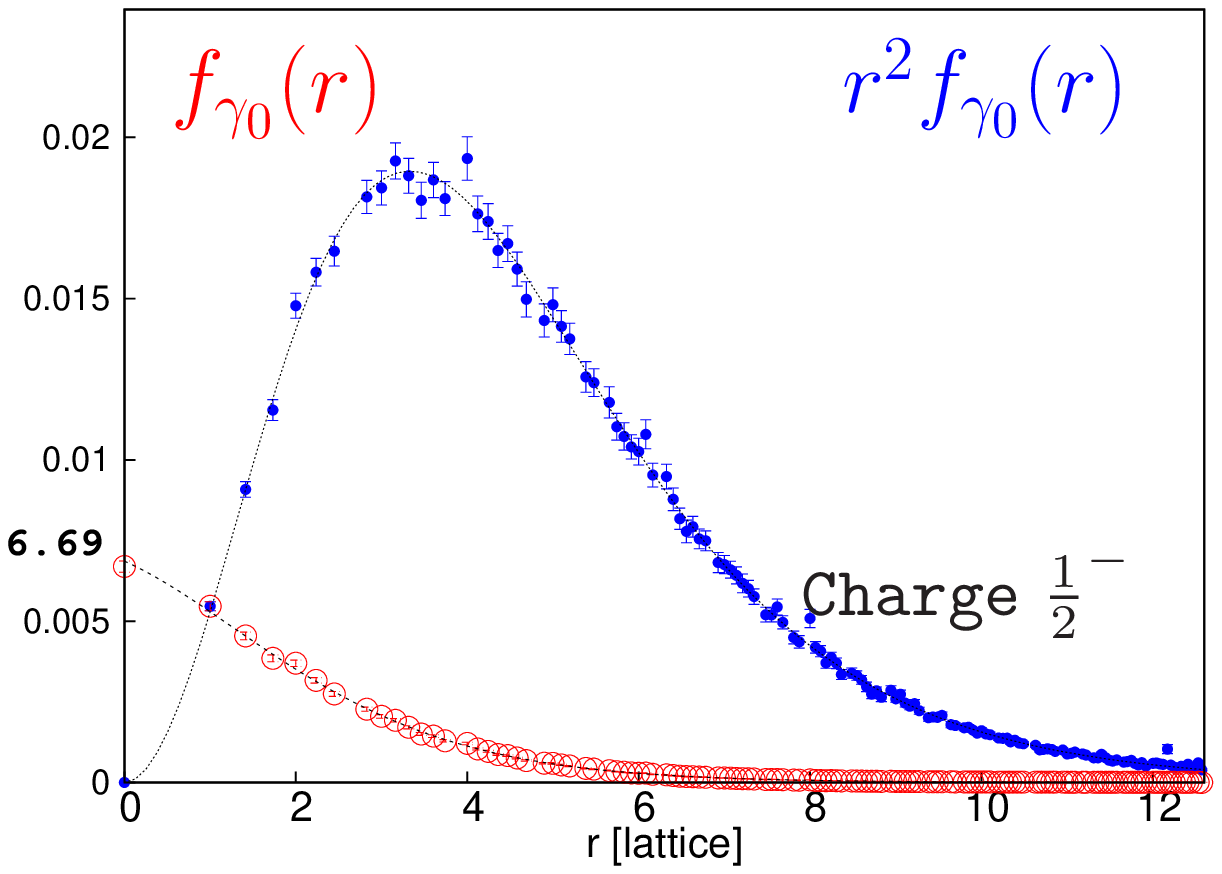}}} & {\resizebox{80mm}{!}{\includegraphics{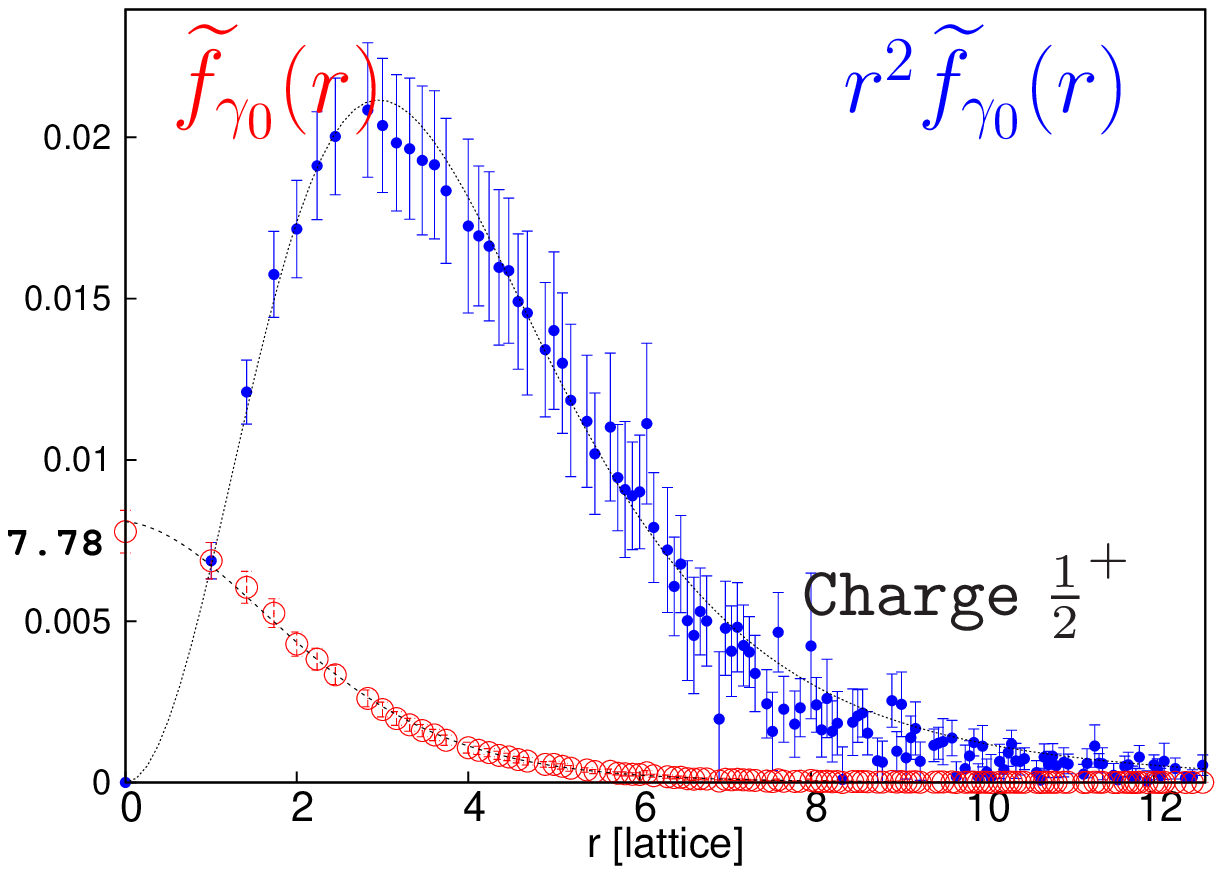}}}\\[-3pt]
{\resizebox{80mm}{!}{\includegraphics{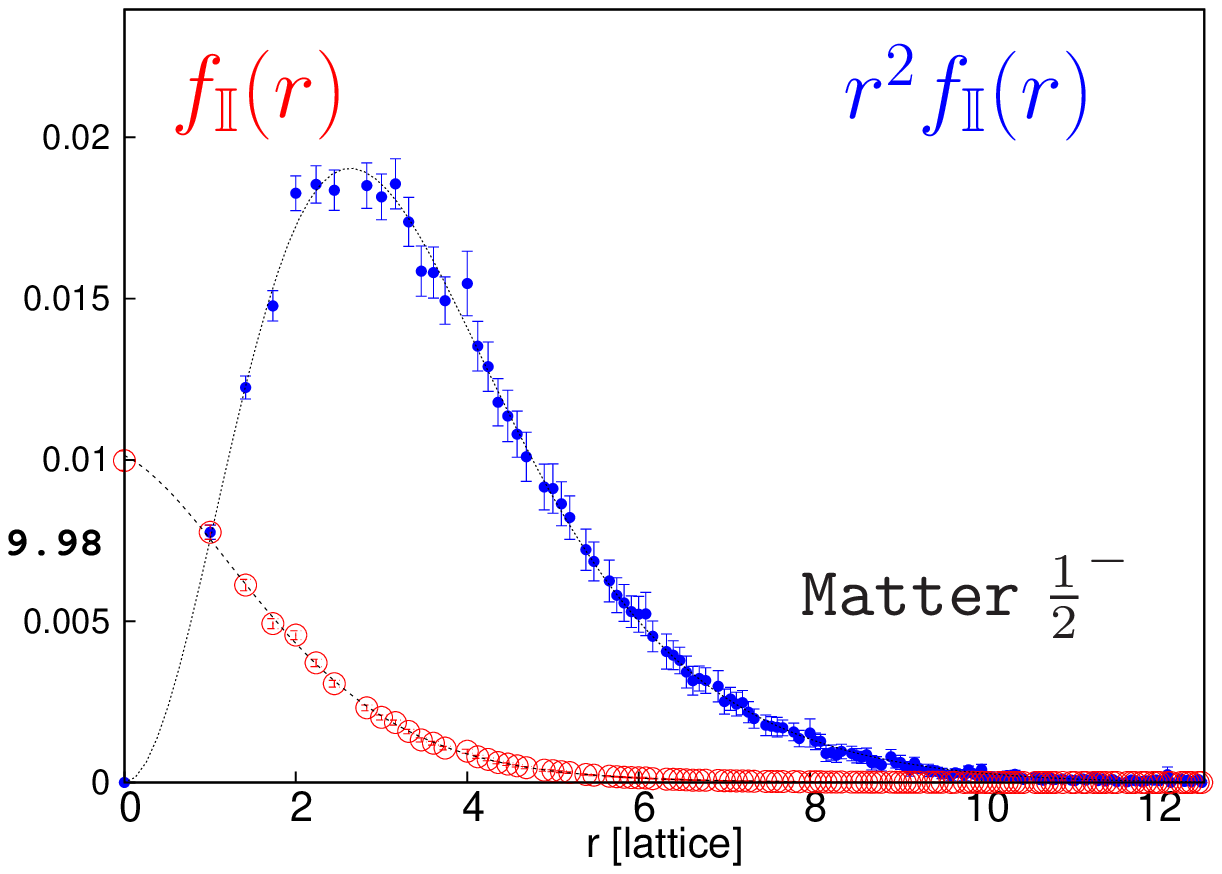}}} & {\resizebox{80mm}{!}{\includegraphics{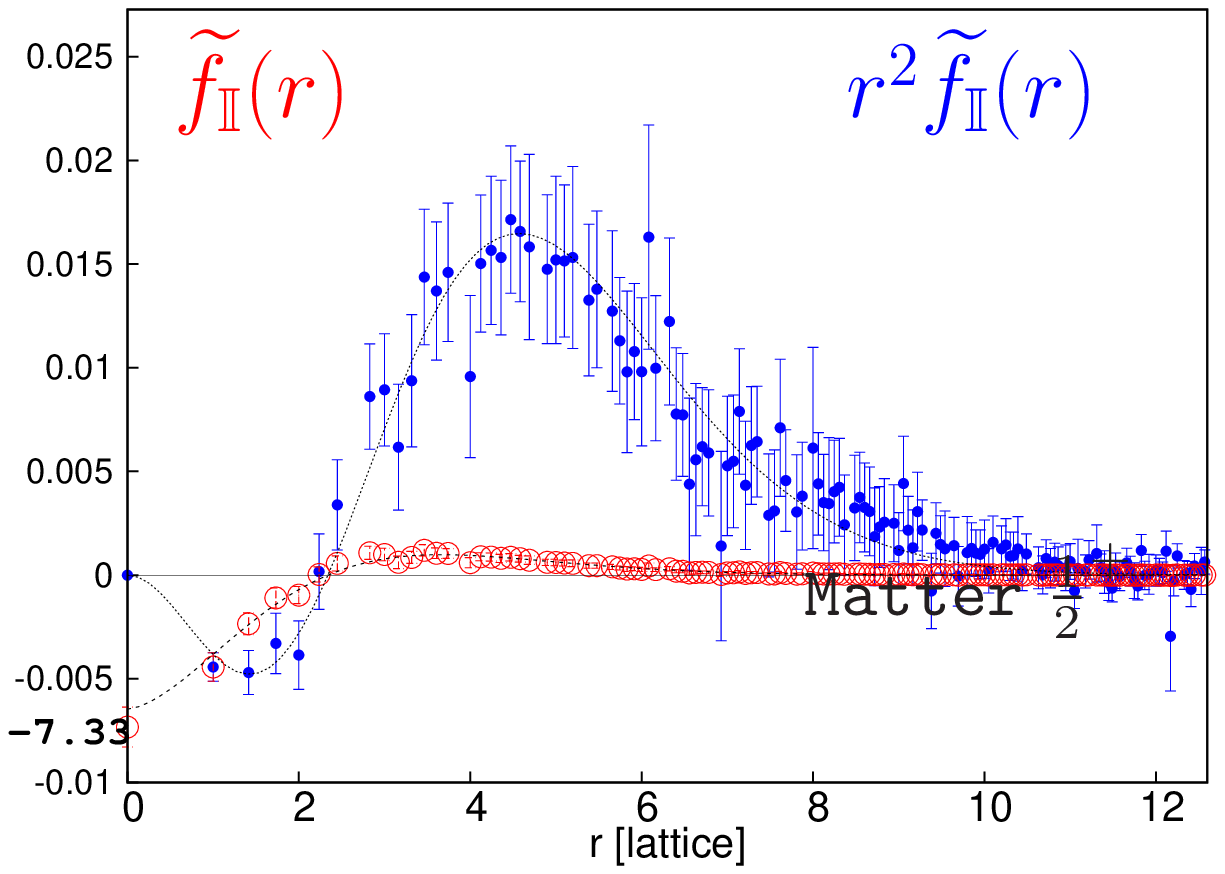}}}\\[-3pt]
{\resizebox{80mm}{!}{\includegraphics{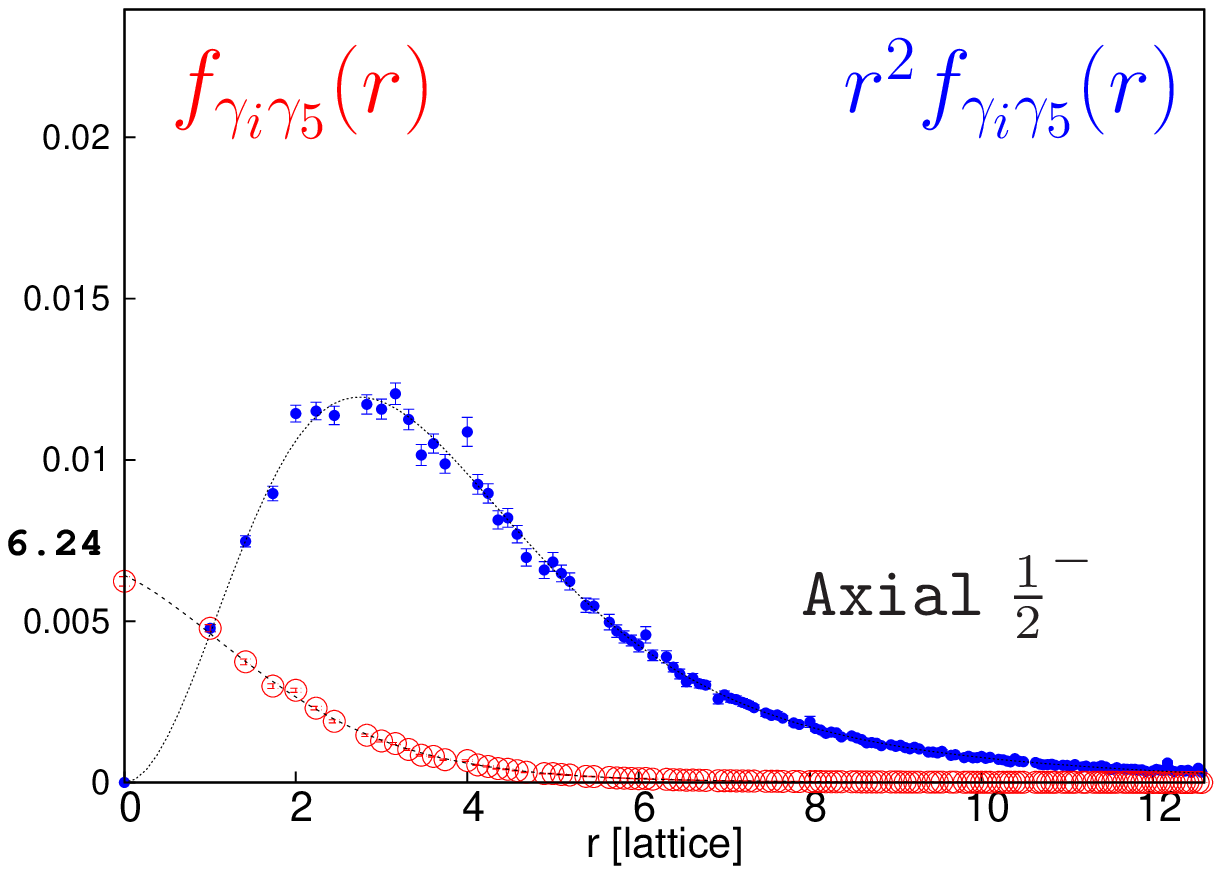}}} & {\resizebox{80mm}{!}{\includegraphics{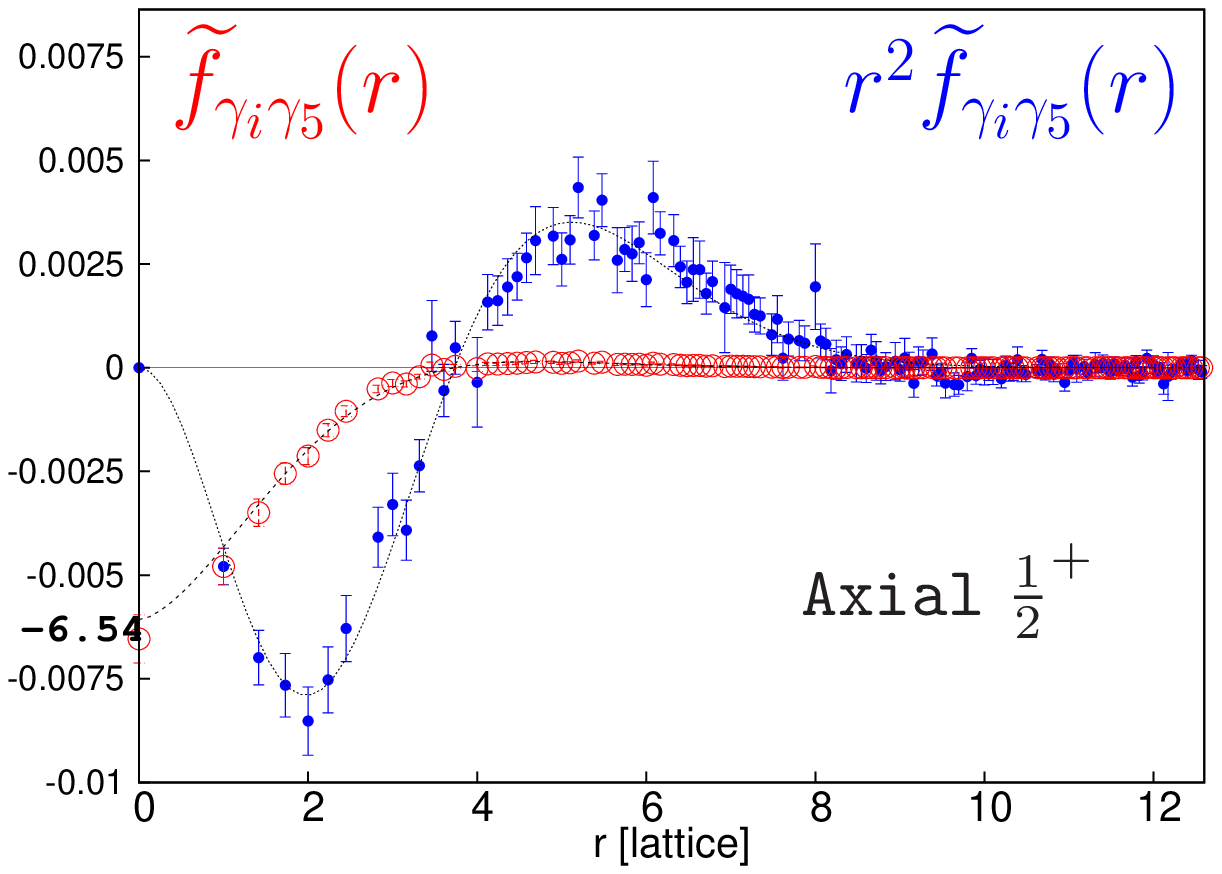}}}
 \end{tabular}
\end{center}
\end{adjustwidth}
\begin{center}
%
%
\caption{\label{fig:2}
\footnotesize{\sl 
Radial distribution of the vector, axial and the scalar density within the static-light meson. On the left side the distributions are obtained
with the external $(1/2)^-$ mesons, and those on the right with the $(1/2)^+$ ones. light-quark is $\kappa_q=0.1374$, which is around the physical
strange quark mass. All results are displayed in lattice units.
}} 
\end{center}
\vspace{-20pt}
\end{figure}

\vspace{-8pt}
\noindent To extract the desired distributions we computed the following three point correlation functions:
%
%
\begin{adjustwidth}{24pt}{0pt}
\vspace{-8pt}
\[
\begin{array}{rrrcll}
\multicolumn{6}{l}{\hspace{-24pt} j^P=(1/2)^- \mbox{doublet, i.e. ground states\, }  J^P=[0^-,1^-], e.g.  (B, B^\ast) }\\[1ex]
C^{(3)}_\Gamma (-t_{s1}, t_{s2}; \vec r ) &=
  {\displaystyle \sum_{\vec x,\vec y}}  \left<  \trout{12pt} \right. \hspace{-3pt} & \bar h(x) \gamma_5 q (x)  \ \bar q(z+\vec r ) & \Gamma & q(z+\vec r) \  \left(\bar h(y) \gamma_5 q (y)\right)^\dagger & \hspace{-4pt} \left. \trout{12pt} \right>_{z - {\rm fixed}},%
  \ \mbox{$\Gamma = \mathbbmss{1}, \gamma_0$} \nonumber \\
C^{(3)}_{\gamma_i \gamma_5} (-t_{s1}, t_{s2}; \vec r ) &=
  {\displaystyle \sum_{\vec x,\vec y}}  \left<  \trout{12pt} \right. \hspace{-3pt} & \bar h(x) \gamma_5 q (x)  \ \bar q(z+\vec r ) & \gamma_i \gamma_5 & q(z+\vec r) \  \left(\bar h(y) \gamma_i q (y)\right)^\dagger & \hspace{-4pt} \left. \trout{12pt} \right>_{z - {\rm fixed}}%
  \nonumber \\[1ex]
\multicolumn{6}{l}{\hspace{-24pt} j^P=(1/2)^+ \mbox{doublet, i.e. first excited states\, }  J^P=[0^+,1^+], e.g. (B_0^\ast, B_1) }\\[1ex]
\widetilde C^{(3)}_\Gamma (-t_{s1}, t_{s2}; \vec r ) &=
  {\displaystyle \sum_{\vec x,\vec y}}  \left<  \trout{12pt} \right. \hspace{-3pt} & \bar h(x) \mathbbmss{1} q (x)  \ \bar q(z+\vec r ) & \Gamma & q(z+\vec r) \  \left(\bar h(y) \mathbbmss{1} q (y)\right)^\dagger & \hspace{-4pt} \left. \trout{12pt} \right>_{z - {\rm fixed}}, %
  \ \mbox{$\Gamma = \mathbbmss{1}, \gamma_0$} \nonumber \\
\widetilde C^{(3)}_{\gamma_i \gamma_5} (-t_{s1}, t_{s2};  r ) &=
  {\displaystyle \sum_{\vec x,\vec y}}  \left<  \trout{12pt} \right. \hspace{-3pt} & \bar h(x) \mathbbmss{1}  q (x)  \ \bar q(z+\vec r ) & \gamma_i \gamma_5 & q(z+\vec r) \  \left(\bar h(y) \gamma_i \gamma_5 q (y)\right)^\dagger & \hspace{-4pt} \left. \trout{12pt} \right>_{z - {\rm fixed}} \nonumber
\end{array}
\]
\end{adjustwidth}
\vspace{-8pt}
The matrix elements are extracted after dividing $C^{(3)}$ by the appropriate two point functions\footnote{%
We take ${Z}\exp(-{E}t)$ to be the analytic form of our two point functions. Due to the heavy quark symmetry the mesons within each doublet are mass degenerate.
and their ${\cal Z}(\widetilde {\cal Z})$'s are equal too.}
\vspace{4pt}
\[
\begin{array}{rclcccccl}
R_{\Gamma}(  \vec r ) &=&
 {\displaystyle C_\Gamma (-t_{s1}, t_{s2}; \vec r ) \over\displaystyle \left( {\cal Z}^{S}\right)^2 \times \exp[- {\cal E} \left( t_{s2}-  t_{s1}\right) ] }\; &
 \stackrel{  |t_{s1}|,t_{s2}\gg 0 }{\xrightarrow{\hspace*{16mm}} }\; &
 \langle B \vert \bar q \Gamma q (\vec r) \vert B\rangle & \equiv & f_{\mathbbmss{1}}(r), f_{\gamma_0}(r) & \mbox{, $\Gamma = \mathbbmss{1}, \gamma_0$} \nonumber\\[-12pt]
&&\multicolumn{2}{l}{
\phantom{
 {\displaystyle C_\Gamma (-t_{s1}, t_{s2}; \vec r ) \over\displaystyle \left( {\cal Z}^{S}\right)^2 \times \exp[- {\cal E} \left( t_{s2}-  t_{s1}\right) ] }\;
 \stackrel{  |t_{s1}|,t_{s2}\gg 0 }{\xrightarrow{\hspace*{16mm}} }\;
}} &
 \langle B \vert \bar q \Gamma q (\vec r) \vert B^*\rangle & \equiv & f_{\gamma_i\gamma_5}(r) & \mbox{, $\Gamma = \gamma_i\gamma_5$} \nonumber \\[-10pt]
\widetilde R_{\Gamma}(  \vec r ) &=&
 {\displaystyle \widetilde C_\Gamma (-t_{s1}, t_{s2}; \vec r ) \over\displaystyle \left( \widetilde {\cal Z}^{S}\right)^2 \times \exp[- \widetilde {\cal E} \left( t_{s2}-  t_{s1}\right) ] }\; &
 \stackrel{  |t_{s1}|,t_{s2}\gg 0 }{\xrightarrow{\hspace*{16mm}} }\; &
 \hspace{-3pt} \langle B_{0}^\ast \vert \bar q \Gamma q (\vec r) \vert B_{1}\rangle & \equiv & \widetilde f_{\mathbbmss{1}}(r), \widetilde f_{\gamma_0}(r) & \mbox{, $\Gamma = \mathbbmss{1}, \gamma_0$} \nonumber\\[-17pt]
&&\multicolumn{2}{l}{
\phantom{
 {\displaystyle \widetilde C_\Gamma (-t_{s1}, t_{s2}; \vec r ) \over\displaystyle \left( \widetilde {\cal Z}^{S}\right)^2 \times \exp[- \widetilde {\cal E} \left( t_{s2}-  t_{s1}\right) ] }\;
 \stackrel{  |t_{s1}|,t_{s2}\gg 0 }{\xrightarrow{\hspace*{16mm}} }\;
}} &
 \langle B_{0}^\ast \vert \bar q \Gamma q (\vec r) \vert B_{1}\rangle & \equiv & \widetilde f_{\gamma_i\gamma_5}(r) & \mbox{, $\Gamma = \gamma_i\gamma_5$} \nonumber
\end{array}
\]
$f_{\mathbbmss{1}}$, $f_{\gamma_0}$, and $f_{\gamma_i\gamma_5}$ are repectively the matter, electric and axial charge distributions for the $(1/2)^-$-states.
The tilde symbol is added to distinguish the distributions for the  $(1/2)^+$- states ($L=1$). The results are presented in figure \ref{fig:2} for the case of
the light quark being of the mass close to that of the physical strange quark. The lattice spacing is about $0.1\ {\rm fm}$.

\newpage
\noindent For more details please see ref.~\cite{structure}. Here we decide to emphasize the following observations:
\vspace{8pt}
\begin{adjustwidth}{-12pt}{0pt}
\begin{itemize}
\item The distributions indeed decrease exponentially when the distance from the static source of color grows.%
\footnote{Except for $\widetilde f_{\mathbbmss{1}}$ and $\widetilde f_{\gamma_i\gamma_5}$, the logarithm of each distributions can be fitted to a polynomial.
$\widetilde f_{\mathbbmss{1}}$ and $\widetilde f_{\gamma_i\gamma_5}$ also die out at large $r$, but to fit them a more elaborate functions are needed, c.f.~\cite{structure}.}
By integrating the distributions over spherical volumes, we see that the cooresponding \emph{charges}, as a function of radius, already saturate at around $1\ {\rm fm}$.
\item We checked that the discrete sum ${\displaystyle \sum_{x,y,z} f_\Gamma (x,y,z)}$\hspace{2pt} and
the integral \hspace{2pt}${\displaystyle 4 \pi \int_0^{\infty} f^{fitted}_\Gamma (r)\ r^2dr}$\hspace{2pt} are in excellent agreement.
This implies that spherical symmetry is well observed.\footnote{Finite lattice spacing only contributes a small correction in this regard.}
The vanishing of the distribution at the boundaries is a necessary condition.
\item Axial and scalar charge distrbutions of the $(1/2)^+$-doublet change sign between $0.2$ and $0.4\ {\rm fm}$. The other distributions are monotonically decreasing.
\item The vector charge distribution remains of the same form when switching the ground state to the excited ones, which is of course consistent with Ward identity.  
As for the axial and scalar charge they are completely different (c.f. fig.~\ref{fig:2}).
\end{itemize}
\end{adjustwidth}
\vspace{-12pt}

\section{$\hat g$ and $\widetilde g$}
$g$ and $\widetilde g$ are the couplings of the soft pion to the mesons within the same doublet  $(1/2)^-$ and $(1/2)^+$, respectively.
Those couplings  are related to the emission/absorption of the $P$-wave soft pion, e.g. $B^\ast \to B\pi$, and $B_1^\prime \to B^{0\ast} \pi$. 
These two axial charges are obtained after integrating the distributions presented in the previous section.  In particular, $g$ is given by 
${\displaystyle \int_0^{\infty}  f_{\gamma_i\gamma_5} (r)\  r^2dr}$  (one can use the discrete sum as definition as discussed above.)
By using the perturbatively determined matching to continuum, our final values are:
\begin{eqnarray}
&&\hat g = 0.526(22)\,, \hspace*{19mm} \widetilde g =-0.157(48)\,,\hspace*{19mm}\textrm{linear chiral extrapolation}\cr
&&\hat g_{\rm HMChPT} =0.444(18)\,, \hspace*{8mm}  \widetilde g_{\rm HMChPT} =-0.14(6)\,,\hspace*{11mm} \textrm{chiral perturbation theory}~\cite{jernej} 
\end{eqnarray}
where we indicate the two chiral extrapolation procedures.  An important remark is that our result clearly shows that the axial charges  $\widetilde g \ll \hat g$. 
This is in contradiction with the claims made on the basis of the chiral quark models of refs.~\cite{doubletrouble} that the dynamics within  $(1/2)^-$ and $(1/2)^+$ doublets are very 
similar~\cite{doubletrouble}. This indeed can be inferred from HMChPT too. However, for that to be the case the charges  $g$ and $\widetilde g$ should be equal, or at least close 
to one another, which is clearly not the case.

Notice that $\hat g$ reported here is in excellent agreement with the recent results presented in~\cite{g-paper} (see also references therein). 
The pionic transition between the heavy-light mesons belonging to different doublets is more complicated a problem, which will be addressed elsewhere.

\section{Charge radius of (Pseudo-)scalar mesons}

Electric charge radius is defined by $\sqrt{< r^2 >}$ where, for the ground state meson,
\[ \left< r^2 \right>_{{\frac{1}{2}}^-} = \left. \int_0^{\infty}  r^2 f_{\gamma_0} (r)\  r^2dr \middle/ \int_0^{\infty}  f_{\gamma_0} (r)\  r^2dr \right. \,, \]
${( \int  f_{\gamma_0} (r)\  r^2dr )^{-1}}$ is simply the Vector Current renormalization constant $Z_V$ obtained non-perturbatively.
The charge radius for excited state is calculated by substituting $f_{\gamma_0}$ by $\widetilde f_{\gamma_0}$.

\vspace{4pt}
\noindent
In the chiral limit, we obtain
$\left< r^2 \right>_{{\frac{1}{2}}^-} = (31.6\pm 0.3)\times a^2\; \Longrightarrow  \left< r^2 \right>^{\rm lin.}_{{\frac{1}{2}}^-} ={\large \mathbf{0.334(3)}}~{\rm fm}^2$.%
\footnote{We have fixed the lattice spacing to  $a=0.0995(4)$~fm, from  $r_0/a=4.695(18)$, as obtained on this same lattice~\cite{CP-PACS}, and by using
$r_0^{\rm phys}=0.467$~fm.  Charge radius of scalar meson is calculated anogolously.}\\
For comparison, the vector meson dominance of the electromagnetic form factor
\begin{eqnarray}
\langle B(p^\prime )\vert J_\mu^{\rm em} \vert B(p)\rangle = F(q^2) \left( p + p^\prime \right)\,, \quad \left< r^2 \right> = 6 \left. {d F(q^2)\over dq^2}\right|_{q^2=0}\,, 
\end{eqnarray}
with $q=p-p^\prime$,  $J_\mu^{\rm em}=e_Q \bar Q\gamma_\mu Q + e_q \bar q\gamma_\mu q$, and $ e_{Q(q)}$ being the electric charge of the
heavy(light)~quark, one obtains that  $ \left< r^2 \right>_{\rm VMD}=6/m_\rho^2 = {\large \mathbf{0.393}}\  {\rm fm}^2$. This result is in good agreement with our result.

\section*{Acknowledgement}
We thank the CP-PACS Collaboration for making their gauge field configurations publicly available, the {\it Centre de Calcul de l'IN2P3 \`a Lyon},
for giving us access to their computing facilities and the partial support of ``Flavianet"  (EU Contract No.~MTRN-CT-2006-035482), and of the ANR
(DIAM Contract  No.~ANR-07-JCJC-0031). We also thank V.~Lubicz for discussions and comments. 

\begin{adjustwidth}{0pt}{-14pt}

\end{adjustwidth}

\end{document}